\documentstyle[preprint,prd,floats,aps,epsf]{revtex}

\begin{document}
\preprint{\vbox{\hbox {December 1998} \hbox{IFP-765-UNC} } }
\title{\bf ADS/CFT String Duality and Conformal Gauge Theories.}
\author{\bf Paul H. Frampton}
\address{Department of Physics and Astronomy,}
\address{University of North Carolina, Chapel Hill, NC  27599-3255}
\maketitle

\begin{abstract}
Compactification of Type IIB superstring on an
$AdS_5 \times S^5/\Gamma$ background leads to SU(N) gauge field
theories with prescribed matter representations. In the 't Hooft
limit of large N such theories are conformally finite. For
finite N and broken supersymmetry ($\cal N$ = 0) I derive the constraints
to be two-loop conformal and examine the consequences for
a wide choice of $\Gamma$ and its embedding $\Gamma \subset 
{\cal C}^3 (\supset S^5)$.
\end{abstract}


\newpage

\bigskip
\bigskip

{\it Introduction.} Recently the relationship of string theory to gauge 
theory received stimulus from the conjecture by Maldacena\cite{mald}
(related earlier papers are \cite{K,GKT,GK,P}) stemming from
string duality which makes in its strongest form the 
assertion that the information contained in
superstring theory is encoded in a four-dimensional gauge field
theory including its non-perturbative sector. This has been
vigorously pursued by many authors, especially Witten\cite{W1,W2,W3}.
A brief review is in \cite{schwarz}.

This relationship appears ironic when one recalls that the
earliest string theories, the dual resonance models for strong interactions, 
were abandoned in favor of an $SU(3)$ gauge theory
25 years ago.  
String theory has generally been 
regarded as much more general than gauge field theory because of its 
far richer structure;
however, that perception was based on perturbative arguments, and the new developments
of Maldacena {\it et al.} are essentially non-perturbative. 

The idea
is to consider N coincident D3-branes with 4-dimensional
world volume theories having superconformal symmetry. This is conjectured 
\cite{mald} to be dual (weak coupling related to strong coupling)
to type IIB superstring theory in a spacetime with
geometry $AdS_5 \times S^5$. The world volume theory is
in this case an ${\cal N}= 4$ supersymmetric Yang-Mills theory
with gauge group $SU(N)$. Originally it is $U(N)$ but this is broken to
$SU(N)$.

The radii of the $AdS_5$ and $S^5$ are equal and both given
by $R = \lambda^{1/4} l_s$ where $\lambda$ is
the\\
't Hooft parameter\cite{tH} $\lambda = g_{YM}^2 N$ ($g_{YM}^2 = g_S$,
the string coupling constant) and $l_S^2 = \alpha^{'}$ the universal
Regge slope. The string tension is $T = (2 \pi \alpha^{'})^{-1}$.

The ${\cal N} = 4$ $ SU(N)$ gauge theory has been known
to be ultra-violet finite for many years\cite{mand}.
This is true not only for $N \rightarrow \infty$, the
conformal limit of Maldacena, but also for finite $N$.

\bigskip
\bigskip

{\it Breaking supersymetries.} By factoring 
out a discrete group $\Gamma$ in $S^5/\Gamma$ it
is possible to break some or all of the ${\cal N} = 4$
supersymmetries. The isometry of $S^5$ is $SO(6) \sim SU(4)$
which may be identified with the R-parity of the ${\cal N} = 4$
conformal gauge theory. The spinors are in the {\bf 4}
and the scalars are in the {\bf 6} of this $SU(4)$. I 
shall here consider only abelian groups $\Gamma = Z_p$,
although non-abelian $\Gamma$ are worth further study
(see {\it e.g.} \cite{HH,GLR}). I 
am considering only $AdS_5 \times S^5/\Gamma$, although
the second 5-dimensional orbifold can be more general
{\it e.g.} the $T^{p,q}$ spaces considered in \cite{KW}.

The number of unbroken symmetries has been studied in
{\it e.g.} \cite{moore,morrison} with the result that
if $\Gamma \subset SU(2)$ there remains ${\cal N} = 2$
supersymmetry; if that is not satisfied but 
$\Gamma \subset SU(3)$ there remains ${\cal N} = 1$
supersymmetry; finally if even that is not
satisfied one is left with ${\cal N} = 0$ or no
supersymmetry. This last case is of most interest here.

It has been demonstrated that the large N limit of
the resultant gauge theory coincides with that
of the ${\cal N} = 4$ case. Such arguments have
been made both using string theory
\cite{vafa} and directly at the field theory
level \cite{bershadsky}. In the latter case
the proof involves a monodromy of the representation
for the group $\Gamma$.

For finite N, however, there is no argument that
the resultant gauge theory is conformal, especially
for ${\cal N} = 0$ where there are no non-renormalization
theorems.

Nevertheless, if there does exist a conformal gauge
theory in four dimensions with ${\cal N} = 0$, it
would be so tightly constrained as to be possibly
unique and would be of interest especially if
it could contain the standard $SU(3) \times SU(2) 
\times U(1)$ model with its peculiar representations
for the quarks and leptons.

The representations which occur in the resultant
${\cal N} \leq 2$ gauge theories from the orbifold construction
have been studied using quiver diagrams\cite{moore}.
I will find that these diagrams, while convenient
for the cases ${\cal N} \geq 1$ need augmentation
for the case ${\cal N} = 0$. 

To specify the potentially conformal gauge theory
I need to state how the group $\Gamma$ is embedded in
${\cal C}^3$. Let the three complex coordinates
of ${\cal C}^3$ be denoted by $\underline{X} = (X_1, X_2, X_3)$.
The action of $Z_p$ is the specified by:
\begin{equation}
\underline{X} \rightarrow (\alpha^{a_1} X_1, \alpha^{a_2} X_2,
\alpha^{a_3} X_3)   \label{Zp}
\end{equation}
where $\alpha = exp(2 \pi i/ p)$ and the three integers
$a_{\mu} = (a_1, a_2, a_3)$ specify the embedding.

In order to ensure an ${\cal N} = 0$ result, I must insist
that $\Gamma$ is not contained in $SU(3)$ by the requirement
that
\begin{equation}
a_1 + a_2 + a_3 \neq 0~~~ (mod~~~ p)   \label{su4}
\end{equation}
At the same time, for the correct behavior of the spinors
we need in addition
\begin{equation}
a_1 + a_2 + a_3 = 0~~~ (mod~~~ 2)    \label{mod2}
\end{equation}
For any given p, there is a finite $\nu(p)$ number of
choices satisfying Eq.(\ref{su4}) and Eq.(\ref{mod2}).
We shall indicate later how to enumerate these $\nu(p)$.

\bigskip
\bigskip

{\it Matter representations.} Because the discrete group
$Z_p$ leads to the identification of $p$ points in ${\cal C}^3$
and the N coinciding D3-branes converge on all $p$ copies, the
gauge group becomes $SU(N)^p$. The surviving states are
invariant under the product of a gauge transformation
and a $Z_p$ transformation defined as in Eq.(\ref{Zp}) above.

For the scalars, it then follows that the scalars fall
into the representations
\begin{equation}
\sum_{\mu} (N_i, \bar{N}_{i \pm a_{\mu}})   \label{scalars}
\end{equation}
For $a_{\mu} \neq 0$ these are bi-fundamentals and for $a_{\mu} = 0$
complex adjoints. If we focus on one $SU(N)$
the only non-singlet representations (the same will be true
for the fermions) are fundamentals, anti-fundamentals and adjoints.
These representations also follow from the Douglas-Moore quiver diagram.

For the fermions we must consider the transformation of a 4-spinor
by making four combinations $A_{\lambda} (1 \leq \lambda \leq 4)$ 
of the $a_{\mu}$

\begin{eqnarray}
A_1 & = & (a_1 + a_2 + a_3) / 2  \\
A_2 & = & (a_1 - a_2 - a_3) / 2  \\
A_3 & = & (-a_1 + a_2 - a_3) / 2   \\
A_4 & = & (-a_1 - a_2 + a_3) / 2 \\
\end{eqnarray}
Again the surviving states are invariant under a product
of the $Z_p$ and gauge transformations. This leads to the fermion representation:
\begin{equation}
\sum_{\lambda} (N_i, \bar{N}_{i+A_{\lambda}}) \label{fermions}
\end{equation}
which can, if required, be deduced from a (different) quiver diagram.

\bigskip
\bigskip

{\it Two-loop $\beta$-functions.} 
I may take the detailed formula for the gauge coupling
$\beta$-function $\beta_g$ from \cite{MV}. The two
leading orders are:
\begin{equation}
\beta_g = \beta_g^{(1)} + \beta_g^{(2)}
\end{equation}
with
\begin{equation}
\beta_g^{(1)} = - \frac{g^3}{(4 \pi)^2} \left[ \frac{11}{3} C_2(G)
- \frac{4}{3} \kappa S_2 (F) - \frac{1}{6} S_2 (S) \right]
\label{beta1}
\end{equation}
and
\begin{equation}
\beta_g^{(2)} = - \frac{g^5}{(4 \pi)^4} \left[ \frac{34}{3} (C_2(G))^2
- \kappa \left[ 4C_2(F) + \frac{20}{3} C_2(G) \right] S_2(F)
-\left[ 2 C_2(S) + \frac{1}{3} C_2(G) \right] S_2(S) + \frac
{2 \kappa Y_4(F)}{g^2} \right]  \label{beta2}
\end{equation}
Here $C_2, S_2$ are the quadratic Casimir, Dynkin index respectively for the
representations indicated, $\kappa$ is 1/2, 1 for Weyl, Dirac fermions
respectively, products like $C_2(R)S_2(R)$ imply 
a sum over irreducible representations and
finally the Yukawa term is included naturally in the two-loop
term (unlike in \cite{MV}) because here
the Yukawa couplings are proportional to the gauge coupling.
The crucial quantity $Y_4(F)$ is defined in terms of the Yukawa matrix
$Y_{ij}^a\psi_i\zeta\psi_j\phi^a$ by
\begin{equation}
Y_4(F) = Tr \left( C_2(F) Y^a Y^{\dagger a} \right)
\label{Y4}
\end{equation}
 
Looking first at ${\cal N} = 4$, the values are easily seen to
$C_2(G)=N, S_2(F)=4N, S_2(S)=6N$ while $C_2(F)S_2(F)=4N^2$ and 
$C_2(S)S_2(S)=6N^2$.
Finally $Y_4(F) = 24g^2N^2$. It follows from Eq.(\ref{beta1}) and 
Eq.(\ref{beta2}) the $\beta_g = 0$ for ${\cal N} = 4$ at two loops,
as is well known\cite{mand}.

However, the situation for ${\cal N} = 0$ is much more complicated.

At one-loop level for ${\cal N} = 0$ the evaluation of $\beta_g^{(1)}$
is the same term-by-term as for ${\cal N} = 4$. This is already
in \cite{KS1,KKS,KS2} for the one-loop
level and since the one-loop $\beta$-function is purely
leading-order in $N$ it conforms to the general arguments
of \cite{vafa,bershadsky}. 

At two-loop order I must examine the non-leading terms in $1/N$
in Eq.(\ref{beta2}). The first, third and fifth terms are always
the same for ${\cal N} = 0$ as for ${\cal N} = 4$, 
respectively $34N^2/3 - 40N^2/3 - 2N^2 = -4N^2$.

To evaluate the second, fourth and sixth terms I find it
necessary to distinguish four cases which are designated 
($\alpha,\beta,\gamma,\delta$) as follows:

\begin{equation}
a_1 = a_2 ; ~~~a_3 = 0. ~~~A_1 = - A_4 \neq 0; ~~~A_2 = A_3 = 0. ~~~~(Case~~~ \alpha). 
\end{equation}
\begin{equation}
a_1 \neq a_2; ~~~a_3 = 0. ~~~A_1 = - A_4 \neq 0; ~~~A_2 = - A_3 \neq 0. ~~~~(Case ~~~\beta).
\end{equation}
\begin{equation}
All ~~~~a_{\mu} \neq 0. ~~~~One ~~A_{\lambda} = 0; ~~~~three ~~A_{\lambda} \neq 0. ~~~~(Case ~~~\gamma).
\end{equation}
\begin{equation}
All ~~~~a_{\mu} \neq 0. ~~~~All ~~A_{\lambda} = 0. ~~~~(Case ~~~\delta).
\end{equation}
These possibilities lead to fermion and scalar representations of $SU(N)$
which are different for the four cases. They exhaust the choices which leave
${\cal N} = 0$ which requires that Eqs.(\ref{su4}),(\ref{mod2}) are fulfilled.
(Note that at least two $a_{\mu}$ must be non-zero).

The evaluation of the remaining terms in 
Eq.(\ref{beta2}) can now be done case by case.
In Case $\alpha$, where both fermions and scalar appear in both
fundamentals and adjoints, we find that 
$C_2(F)S_2(F) = 4N^2(1 - 1/(2N^2))$,
$C_2(S)S_2(S) = 6N^2(1 - 2/(3N^2))$ 
and $Y_4(F) = (24N^2 - 16)g^2$.
Substituting in Eq.(\ref{beta2}) leads, as generally expected to
an non-vanishing $\beta_g$ for finite N and a non-conformal
gauge theory.

For the other cases, I find for the three group theory
expressions 
$C_2(F)S_2(F)$, $C_2(S)S_2(S)$ and $Y_4(F)$ respectively
the following:
\begin{equation}
4N^2(1 - 1/(N^2)), ~~~6N^2(1 - 2/(3N^2)) ~~~and~~~(24N^2 - 24). 
~~~(Case~~~\beta)
\end{equation}
\begin{equation}
4N^2(1 - 3/(4N^2)), ~~~6N^2(1 - 1/(N^2)) ~~~and~~~(24N^2 - 18). 
~~~(Case~~~\gamma)
\end{equation}
\begin{equation}
4N^2(1 - 1/(N^2)), ~~~6N^2(1 - 1/(N^2)) ~~~and~~~(24N^2 - 24). 
~~~(Case~~~\delta)
\end{equation}
Substituting in Eq. (\ref{beta2}), I find that $\beta_g^{(2)}$
is non-vanishing except in the Case $\gamma$. For this surviving
theory, the
fermions are in both fundamentals and adjoints, while all scalars 
are in fundamentals. This is therefore the only combination of 
matter representations of further interest.

\bigskip
\bigskip

{\it Directions.} A subsequent question to be addressed is what happens at 
three-loop and even higher orders. Also one must consider running
of the Yukawa and quartic Higgs self-couplings due to 
possible non-vanishing of their $\beta$-functions
$\beta_Y$ and $\beta_H$. It is planned to publish a more 
complete analysis elsewhere;
I conclude this proposal with comments and possible future directions. 

Often low-energy supersymmetry is adopted in order
to solve the hierarchy problem
of the Planck or GUT scale to the weak
scale. This hierarchy is theory-generated and
one may instead be agnostic about physics at 
$\gtrsim 1000$TeV scale where there is no real information.
For example, recent ideas about extra Kaluza-Klein dimensions
at reduced scales {\it e.g.} \cite{DDG1,DDG2,DDG3,ADD}
avoid the hierarchy altogether and hence 
remove the main motivation for 
low-energy supersymmetry.

The possible role of an ${\cal N} = 0$, $d=4$ conformal gauge 
theory may be put in 
context by imagining 
the level of skepticism to infinite renormalization of QED in 1948 
(and later of the standard model) if the 
example of \cite{mand} had been found four decades earlier.

The exciting possibility is that the standard model is
part of such an ${\cal N} = 0$ conformal gauge theory.
The mass scales $\Lambda_{QCD}$ and $M_W$
would arise from necessarily non-perturbative 
effects, and gravity would
be accommodated through the holographic principle\cite{tH2,W1}.
Using AdS/CFT duality could help identify the relevant
conformal theory. If so, this could shed light on
the outstanding questions (families, CP violation, etc.)
posed by the standard model.

\bigskip
\bigskip
\bigskip
\bigskip
\bigskip
\bigskip
\bigskip 

\newpage

I wish to acknowledge useful discussions with
J. Lykken, D.R. Morrison and S. Trivedi.
This work was supported in part by the
US Department of Energy under Grant No. 
DE-FG05-85ER-40219.

\end{document}